% TEX SET-UP 
% TO BE PUBLISHED IN THE BULLETIN OF ASTRONOMICAL SOCIETY OF INDIA
% Please do not change the page layout, except \hoffset and/or
% \voffset, where changes may be required depending on default printer
% settings. Leave a margin of 3.8 cm on the left and 5 cm at the top

\magnification=\magstep0
\hsize=13.5 cm               %  horizontal size of printed page
\vsize=19.0 cm               %  vertical size of printed page
\baselineskip=12 pt plus 1 pt minus 1 pt  % The line spacing
\parindent=0.5 cm  % The paragraph indent
\hoffset=1.3 cm      % The horizontal offset (may need to be changed)
\voffset=2.5 cm      % The vertical offset (may need to be changed)
\font\twelvebf=cmbx10 at 12truept % Set bold font for Title
\font\twelverm=cmr10 at 12truept % Set large font for Name
\overfullrule=0pt
\nopagenumbers    %  Actual page nos. will be inserted by the Editor
%
% The headlines
% The changes in the headlines should be made just before the Abstract
\newtoks\leftheadline \leftheadline={\hfill {\eightit G.P. Rowell}
\hfill}
\newtoks\rightheadline \rightheadline={\hfill {\eightit SNRs \& TeV Gamma-Ray Astronomy}
 \hfill}
% Do not change the headline on the first page of paper.
\newtoks\firstheadline \firstheadline={{\eightrm Bull. Astron. Soc.
India (1998) {\eightbf xx,} } \hfill}
\def\makeheadline{\vbox to 0pt{\vskip -22.5pt
\line{\vbox to 8.5 pt{}\ifnum\pageno=1\the\firstheadline\else%
\ifodd\pageno\the\rightheadline\else%
\the\leftheadline\fi\fi}\vss}\nointerlineskip}
%
% Defining 8-pt fonts for figure captions and references
\font\eightrm=cmr8  \font\eighti=cmmi8  \font\eightsy=cmsy8
\font\eightbf=cmbx8 \font\eighttt=cmtt8 \font\eightit=cmti8
\font\eightsl=cmsl8
\font\sixrm=cmr6    \font\sixi=cmmi6    \font\sixsy=cmsy6
\font\sixbf=cmbx6
%
%for switching to eight point type \eightpoint
\def\eightpoint{\def\rm{\fam0\eightrm}
\textfont0=\eightrm \scriptfont0=\sixrm \scriptscriptfont0=\fiverm
\textfont1=\eighti  \scriptfont1=\sixi  \scriptscriptfont1=\fivei
\textfont2=\eightsy \scriptfont2=\sixsy \scriptscriptfont2=\fivesy
\textfont3=\tenex   \scriptfont3=\tenex \scriptscriptfont3=\tenex
\textfont\itfam=\eightit  \def\it{\fam\itfam\eightit}%
\textfont\slfam=\eightsl  \def\sl{\fam\slfam\eightsl}%
\textfont\ttfam=\eighttt  \def\tt{\fam\ttfam\eighttt}%
\textfont\bffam=\eightbf  \scriptfont\bffam=\sixbf
\scriptscriptfont\bffam=\fivebf \def\bf{\fam\bffam\eightbf}%
\normalbaselineskip=10pt plus 0.1 pt minus 0.1 pt
\normalbaselines
\abovedisplayskip=10pt plus 2.4pt minus 7pt
\belowdisplayskip=10pt plus 2.4pt minus 7pt
\belowdisplayshortskip=5.6pt plus 2.4pt minus 3.2pt \rm}
%
% define the displayed equations to be indented 1.5 cm from left
% as required by the Bulletin. Hopefully this will work for all
% equations. With this definition using the normal $$...$$ should
% produce the equations with correct indentation, but it will be necessary
% to use \eqno to put equation numbers (though it will be possible to put
% blank eq. nos.). Further, eq. nos using \eqalignno will not work
%
\def\leftdisplay#1\eqno#2$${\line{\indent\indent\indent%
$\displaystyle{#1}$\hfil #2}$$}
\everydisplay{\leftdisplay}
%
% Some useful definitions
% less than or order of \la
\def\frac#1#2{{#1\over#2}}
\def\la{\mathrel{\mathchoice {\vcenter{\offinterlineskip\halign{\hfil
$\displaystyle##$\hfil\cr<\cr\sim\cr}}}
{\vcenter{\offinterlineskip\halign{\hfil$\textstyle##$\hfil\cr<\cr\sim\cr}}}
{\vcenter{\offinterlineskip\halign{\hfil$\scriptstyle##$\hfil\cr<\cr\sim\cr}}}
{\vcenter{\offinterlineskip\halign{\hfil$\scriptscriptstyle##$\hfil\cr<\cr%
\sim\cr}}}}}
% greater than or order of \ga

%
%
%to generate boldface characters
\def\pmb#1{\setbox0=\hbox{$#1$}\kern-0.015em\copy0\kern-\wd0%
\kern0.03em\copy0\kern-\wd0\kern-0.015em\raise0.03em\box0}
%
%Beginning of Document%
\pageno=1
\vglue 60 pt  %Leave some space on page 1 before the title
% The title
%
\leftline{\twelvebf  Cosmic Rays from SNRs and TeV Gamma-Ray Astronomy} 
% if more than one line is required for the title, then use next two lines ...
%
\smallskip
% end of title
\vskip 46 pt  % Space between title and author(s) name(s).
\leftline{\twelverm G.P. Rowell} % Name of Authors
\vskip 4 pt
\leftline{\eightit Max Planck Institut f\"ur Kernphysik, D-69029 Heidelberg}
%\leftline{\eightit  to reduce the number of lines}
%
% If authors are from different institutes, repeat the above lines
% for each institution. For authors from same institution write the
% names in one line.
%
%\vskip 0.5 cm
%\leftline{\twelverm V. R. Co-author1 and V. R. Co-author2}
%\vskip 4 pt
%\leftline{\eightit Name and Address of the institution}
\vskip 20 pt % leave some space between author(s) names(s) and abstract
%
%
% The leftheadline should include the Authors' name, for two authors use
% \&  (e.g. I. M. Author \& I. M. Co-author) for three or more authors
% use et al.,
\leftheadline={\hfill {\eightit G.P. Rowell} \hfill}
% Use a short running title as the rightheadline
\rightheadline={\hfill {\eightit Cosmic Rays from SNRs and TeV Gamma-Ray Astronomy}  \hfill}

% Abstract begins
%
{\parindent=0cm\leftskip=1.5 cm

{\bf Abstract.}
\noindent The origin of Galactic cosmic rays is still a burning question that 
forms a major motivation for developments in ground-based gamma-ray astronomy. 
SNRs are long-thought to be sites for the acceleration of Galactic cosmic rays,
and evidence for gamma-ray and non-thermal X-ray production from some SNRs suggest 
that they may be capable 
of accelerating particles to multi-TeV energies. Yet, along with this, and in the 
same overall model framework 
(diffusive shock acceleration), is the need to accommodate upper limits at TeV energies imposed 
on other examples.  This review will present an update on the status of 
SNR observations at TeV energies, their interpretation, and
%(2) Pose the question. 
%"Are we able to at present state definitively whether SNRs as an object class 
 discuss the relevant parameters and issues of next generation ground-based instruments 
relating to their ability to confirm SNRs as Galactic cosmic ray sources.           
       
\smallskip 
\vskip 0.5 cm  %  Space between Abstract and Key words
{\it Key words:}  Supernova remnants, gamma-rays,  

}                                 %  End of abstract
% Beginning of document
%
%
% Beginning of a section heading
%
% for the first section leave 20 pt space, for subsequent sections just
% leave bigskip (i.e. 12 pt)
\vskip 20 pt
\centerline{\bf 1. Introduction}
\bigskip
\noindent
Supernova remnants (SNR), notably of the shell type, have long been 
considered the most likely accelerators of Galactic cosmic rays (CR) up to
at least 10$^{14}$ eV and possibly the knee (E$\sim10^{15}$ eV). 
SNRs as a collective are able to meet the energetics of the observed CR flux at
Earth, when considering CR escape and SNR birthrate issues. The average 
mechanical energy released by each SNR is $\sim10^{51}$ erg, and in the above context
about 10\% of this energy is required for CR acceleration to relativistic energies.  
The diffusive shock acceleration process invoked in SNR also naturally explains 
the dN/d$E\sim E^{-2.0}$ spectrum obtained for the observed CRs at Earth after correction of 
propagation effects (ie. the source spectrum). 
The advent of sensitive space-based gamma-ray and X-ray detectors in recent years has 
provided the opportunity to identify likely sites of CR acceleration in our galaxy. 
(eg. Esposito et al. 1996, Slane 2001). However,  in the context of our
understanding of radiative processes from relativistic particles,
the search for gamma-radiation at multi-GeV to TeV energies (or Very High
Energies, VHE) is also deemed vital to this effort.
Within this energy regime it is the atmospheric \v{C}erenkov imaging 
technique that appears to offer the best sensitivity to detect gamma radiation of such 
energies.

In this review I will present a summary of shell-type SNR observations at TeV energies, and 
then cover relevant issues for the next generation ground-based instruments now under 
construction to confirm or deny the theory that these objects are responsible for 
accelerating CRs.

%We 
%will begin with a brief summary of our current understanding of known sources of
%TeV gamma radiation is a good place to lead into the main theme. 

\vskip 20 pt
\centerline{ {\bf  2.   VHE Gamma Ray Astronomy and \v{C}erenkov Imaging}    }
\bigskip 
\noindent

Non-thermal high energy radiation is presently the most accessible tracer of
cosmic ray acceleration in the universe, by virtue of the physics 
associated with relativistic particles and high energy photons. The spectra of such
radiations also closely reflects the spectra of parent particles, and so one
is able to study particle acceleration processes. 

The detection of VHE gamma-rays uses ground-based sampling of the extensive air shower
(EAS). EAS comprise the secondary particles generated as primary gamma and 
cosmic rays interact with the Earth's atmosphere.
The \v{C}erenkov  signature of EAS carries information about the primary particle's 
direction, energy and nature (hadronic or electromagnetic).
Viewing the {\it angular distribution} of this signature with a sufficiently large (usually segmented) 
mirror ($\geq$ 10 m$^2$) and focal plane array or camera of 
photomultiplier (PMT) pixels (ie. an imaging atmospheric \v{C}erenkov telescope, IACT) allows to
reconstruct primary parameters to high accuracy with a remarkably high 
effective collection area ($\ge10^5$ m$^2$). Note that a number of experiments
today are also making use of the {\it lateral distribution} to detect gamma-rays,
a subject which is left at the moment for later discussion.
The first reliable detection of 
the Crab Nebula by the Whipple Collaboration with the imaging technique
(Weekes et al. 1989) was made using a single telescope employing a 75 m$^2$ 
segmented mirror and 37 PMT camera. Since then, improvements have been realised 
such that cameras in use today by Whipple 
(Finley et al. 1999), CAT (Barrau et al 1998), HEGRA (Daum et al 1997), 
CANGAROO II (Yoshikoshi et al. 1999) and TACTIC (Bhat et al. 2000) now all exceed 200 
pixels and achieve pixel sizes better than $\sim$0.25$^\circ$.
EAS \v{C}erenkov images are often parameterised by the moments of an ellipse 
following Hillas (1985) although some improvements have been demonstrated 
(eg. LeBohec et al. 2000, Akhperjanian \& Sahakian 1999). The primary aim is to 
preferentially select EAS images of a gamma-ray nature against those of the vastly
outnumbering CR background, leaving a statistically significant excess of gamma-ray 
like events.  A further significant improvement in the 
technique has been the use of stereoscopic imaging in which the EAS is viewed
by at least two different telescopes separated by $\sim$100 m. This  
takes advantage of the uncorrelated nature of EAS image fluctuations,
thereby achieving an improvement in angular resolution roughly proportional 
to $1/\sqrt{n}$ with $n$ the number of views attained for each EAS image
(Hofmann et al. 1999).  The HEGRA CT-System is currently the most sensitive example of such
an array employing the stereoscopic technique. Today angular, energy resolution, energy threshold, 
and sensitivity of $\la 0.1^\circ$, $\sim$15\%, $\sim$250 GeV and 
$\sim1\times10^{-12}$ erg cm$^{-2}$ s$^{-1}$ ($>$1 TeV, 50 hours, 
5$\sigma$) respectively are achieved by the best ground-based systems. 
Here we define the energy threshold at that energy which maximises the differential
trigger rate for gamma-rays, for a variety of source spectra. The achieved sensitivity
is equivalent to $\sim$1 Crab at 5 $\sigma$ significance, and has been sufficient to 
probe the TeV-output
of the two BL-Lac blazars Mrk 421 and Mrk 501 on timescales of tens of {\it minutes} during flaring 
episodes.
%The advantage of a large collecting area and good sensitivity area offered by IACTs is immediately
%apparent after comparison of results from IACTs and EGRET on 
For further reviews of TeV observations and current 
instrumentation the reader is directed to Weekes (2000) and Fegan (2001).

\vskip 20 pt
\centerline{  {\bf 3.  Cosmic Ray Acceleration in SNRs}    }
\bigskip
\noindent

The considerable observational evidence at X-ray energies that SNRs are 
capable of accelerating CR electrons to multi-TeV energies, arises from the interpretation
that their, in some cases entirely featureless X-ray spectra result 
from synchrotron emitting electrons (Koyama et al. 1995, Koyama et al. 1997, Slane et al. 2001). 
One can then make rather straightforward predictions concerning the GeV/TeV emission from such
sources. The GeV/TeV emission arises from inverse Compton boosting of ambient photons 
(usually the microwave background) by the X-ray synchrotron-emitting high energy electrons
(synchrotron/inverse Compton or S/IC theory), 
and that the ratio of the X-ray to gamma-ray flux is related to the shock-compressed magnetic
field $f_{\rm x}/f_{\rm \gamma}\sim B^2$. Observations at TeV energies allow to
determine and/or place constraints on $B$, which is an important parameter determining many features 
of shock acceleration. TeV observations can validate the S/IC framework since alternative 
explanations have been suggested for the hard X-ray emission in some cases (Laming 1998, Laming 2001).
Recent results from CANGAROO suggesting the two SNRs, SN1006
and RXJ1713.7$-$3946 are TeV gamma-ray emitters (Tanimori et al. 1998, 
Muraishi et al. 2000), if confirmed could be considered strong evidence for the 
S/IC theory, although the hadronic channel (discussed shortly) may not be ruled out for SN1006
(Aharonian \& Atoyan 1999).

The widely-accepted framework of diffusive shock acceleration that provides
acceleration for such electrons however, must also do so for hadrons
(Blandford \& Eichler 1987, V\"olk 2001). 
Evidence for hadron acceleration in SNRs is unfortunately rather inconclusive in the face 
of that for electron acceleration. The predictions of Drury, Aharonian \& V\"olk (1994)
hereafter DAV, and also Naito \& Takahara (1994) led to intense observation of SNRs by space 
and ground-based gamma-ray observatories.  DAV showed 
that an appreciable gamma-ray flux may be obtained from the decay of $\pi^\circ$ particles 
produced in the interaction of shock-accelerated hadrons with ambient matter nearby, ie. the 
so-called hadronic channel, which scales according to $F_\gamma \sim E_{\rm cr}\,n\,/d^2$, for the
SNR distance $d$, ambient matter density $n$, and energy available for particle acceleration 
$E_{\rm cr}$ (canonically believed to be $\sim$10\% the total kinetic SNR energy). $F_\gamma$  is also
somewhat dependent on the spectral index of parent particles. Such fluxes 
were marginally close to sensitivities of instruments operating throughout the 1990's, when considering 
plausible values for the parameters just mentioned. However, apart from the
results of EGRET (Esposito et al. 1996) which showed that the MeV/GeV emission from some likely
SNRs candidates is consistent with the idea that they are areas
of enhanced CR density, results from ground-based efforts have generally revealed TeV upper limits
at levels of order 10\% Crab flux.
The exceptions have been recent evidence that Cas-A is an emitter of 
TeV gamma-rays (Aharonian et al. 2001a) and of course the two aforementioned CANGAROO results. 

Throughout the last half-decade, over a dozen SNRs have undergone observation at TeV energies. Table 1 
summarises results for some of these, and includes information considered
indicators of the likelihood of TeV emission. We can see that a reasonable range of SNR ages
and environments has been sampled although most are of the shell-type. Those SNRs exhibiting such properties 
as proximity, correct age (explained below), non-thermal X-ray emission, an EGRET source association
and possible 
interaction with an adjacent molecular cloud (with high $n$) would be considered prominent 
candidates for TeV gamma-ray emission.

\bigskip
\vbox{\tabskip=7.5pt 
\hrule \smallskip \hrule \smallskip
%\settabs 8 \columns
\halign to \hsize{#\hfil&\hfil#&\hfil#&#\hfil&\hfil#\hfil&\hfil#\hfil&\hfil#\hfil&\hfil#\hfil&\hfil#
      \hfil&\hfil#\hfil \cr
SNR & radius      & age$^{a}$ & dist  & SNR  & TeV & EGRET     & Non-th & SN   & Molec. \cr
    & [$^\prime$] &  [kyr]    & [kpc] & type &     &           & X-rays & type & cloud  \cr 
\noalign{\smallskip \hrule \smallskip} 
W28          &    30 & 150  & 2.3 & C  & N & Y & N & II? & Y \cr   
W44          & 35x27 &  20  & 4.0 & C  & N & Y & N & II? & Y \cr  
W51          &    30 &  30  & 6.0 & S? & N & M$^{c}$ & N & II? & Y \cr   
W63          & 95x65 &  24  & 2.0 & S  & N & M & N & II? & Y \cr   
$\gamma$-Cyg &    60 &  10  & 1.8 & S  & N & Y & N & II? & M \cr   
Monoceros    &   220 &  10  & 1.6 & S  & M & Y & N & II? & Y \cr  
Tycho        &     8 & 0.4  & 2.8 & S  & N & N & M & Ia & M \cr   
IC443        &    45 & 6.2  & 2.0 & S  & N & Y & N & II? & Y \cr   
Cas-A        &     4 & 0.3  & 3.4 & S  & Y & N & Y & Ib & M \cr   
SN1006       &    30 & 1.0  & 1.8 & S  & Y & N & Y & Ia & N \cr   
RXJ1713.7    & 65x55 & 40.0 & 6.0 & S? & Y & N & Y & ?  & Y \cr                                      
\noalign{\smallskip \hrule \smallskip \hrule \smallskip}}
{\eightpoint a: Upper Limits \hskip 0.5cm b: $\sim$50\% uncertainty \hskip 0.5cm c: M=Maybe!}
\bigskip
{\eightpoint 
\noindent {\bf Table 1.} The present status of prominent SNRs observed at 
      X-ray to TeV gamma-ray energies with the column 'TeV' indicating a detection at
      TeV energies (Y: Yes, reported by least one research group) or upper limit only (N: No) at typically
      $\sim$10\% Crab flux. The presence of non-thermal
      X-ray, EGRET (MeV/GeV) emission, and whether an interaction with a molecular
      cloud is suspected (high $n$, and OH maser emission) are important indicators of 
      TeV emissivity. Also 
      included is the SNR type (C-Composite, S-Shell) and supernova (SN) progenitor type, 
      if known. References
      for each source are as follows; W28 (Rowell et al. 2000); W44, W51, W63, 
      $\gamma$-Cyg, IC443 (Buckley et al. 1998); Monoceros (Lessard et al. 1999,
      Lucareli et al. 2000); Tycho (Aharonian et al. 2001b); Cas-A (Aharonian et al.
      2001a); SN1006 (Tanimori et al. 1998); RXJ1713.7 (Muraishi et al. 2000). Where possible,
      radius, age and distance information are taken from Green (2000).}
\bigskip}
Attempts to explain these results overall, usually in
combination with those at X-ray and EGRET gamma-ray energies are not trivial, involving many
issues. In the context of accommodating upper limits, extrapolations from 
EGRET up to TeV energies when considering just the hadronic channel, appear to often imply 
rather steep particle spectra, $\sim E^{-2.4}$ (Buckley et al. 1998). More than one process may be 
responsible for the EGRET emission in some cases. For example Gaisser et al. (1998) and Sturner (1997) 
showed that the low energy gamma-ray emission may result from non-thermal electron Bremsstrahlung, again
with rather steep particle spectra (together the 
IC and non-thermal Bremsstrahlung comprise the electronic channel of TeV gamma-ray production). 
Time evolution of SNR gamma-ray 
emissivity must also be considered, as DAV showed that the peak gamma-ray emissivity of a SNR  
occurs during the so-called Sedov phase, corresponding to the time after which the swept-up 
mass exceeds that ejected by the blast. Later the non-linear model of Berezkho \& V\"olk (1997) also
confirmed this profile, albeit with some differences. Such information is likely important for those 
examples considered borderline Sedov (eg. Tycho's SNR, Aharonian et al. 2001b), or even too old where age 
limitations on maximum particle energies come into play (see Drury et al. 2000). Non-linear theory 
(Baring et al. 1999) also indicates that
SNR expansion into high density environments may actually limit the maximum particle energies 
to $\le$1 TeV,
thereby anti-biasing SNR selection according to $n$. Complications are expected from 
SNR expanding into wind bubbles as expected from type Ib and II supernovae 
with massive progenitors (Berezkho \& V\"olk 2000). 
And finally, aside from the above issues and those concerning shock acceleration itself
(Kirk \& Dendy 2001, Drury et al. 2001), we have also to deal with the rather large 
uncertainties (factors of least 2) in observable parameters such as 
$d$, $n$ and $E_{\rm cr}$ contributing to the wide parameter space, often a factor $\sim10$, in which 
to accommodate models. 

\midinsert
%\epsfxsize=0.8\hsize
%\hskip .5 in
\vskip 8cm
\includegraphics{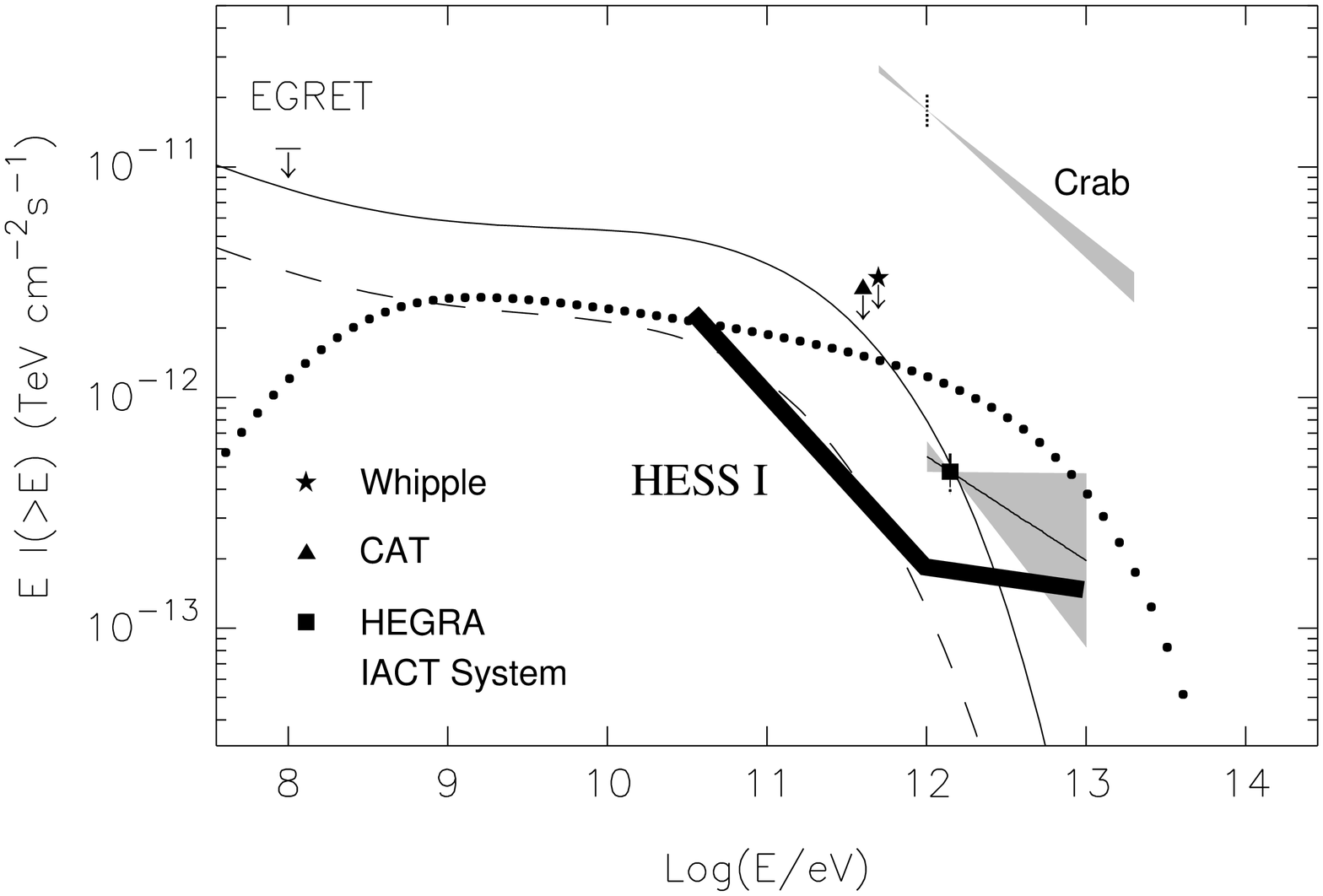} 
      
{\eightpoint {\bf Figure 1.} HEGRA Cas-A flux and 1$\sigma$ uncertainty on spectral index compared 
with the modelling of Atoyan et al. (2000). Hadronic (dotted) and electronic (IC+Bremsstrahlung, 
solid and dashed) channels are indicated. Reasonable scaling parameters for each channel were assumed. 
For comparison the Crab flux and error on spectral index are included. The Cas-A flux amounts to 
3.3\% Crab flux.
The Whipple (Lessard et al. 1999) and CAT (Goret et al. 1999) upper limits are indicated. 
Note the estimated sensitivity of H.E.S.S. Phase I for point sources (5$\sigma$ detection at 50 hrs 
observation shaded region), which will be more or less similar to that of VERITAS and CANGAROO III.}
\endinsert
The results of Cas-A observations nicely illustrate the present situation
(Figure 1). Deep observations of Cas-A (over 200 hrs) by the HEGRA CT-System have revealed a 
TeV gamma-ray flux at a level $\sim$3\% that of the Crab (Aharonian et al. 2001a). 
Atoyan et al. (2000) used a tri-zone model to characterise areas of different magnetic field and
electron transport in the range 0.1 to 1 mG for the electronic channel (IC + Bremsstrahlung),
and a total energy of 2$\times 10^{49}$ erg in CR hadrons. 
%Despite the complexity of the model,
%the rather high magnetic field ensures that the electron cut-offs are well below that for hadrons,
%predicting spectral differences at $E \ge 1$ TeV.
Here, we can clearly see the limitations of present-day ground-based instruments in sampling such a flux. 
Owing to the dependence
of sensitivity with $\sqrt{t}$ it is unlikely that further observations of Cas-A will be carried out.
Most importantly, insufficient statistics prevent an accurate estimate
of the spectral index, and hence discrimination between hadronic and electronic TeV channels is 
currently not available. 
Thus it is apparent that detailed studies of sources with fluxes around 10\% Crab will only be accessible
after significant improvements particularly in sensitivity. A reduction in energy threshold 
below the $\sim$250 GeV available today is also a strong motivation, as it would achieve a large 
dynamic spectral range, allow studies of SNR with lower energy cutoffs, and of course help in
detection of the EGRET unidentified sources. 
%The latter point is important in cases where limitations
%on maximum particle energies $\le 10$ TeV are imposed.

Overall, since it appears that many TeV upper limits lie not far from the conservative edge of parameter 
space, it is generally accepted that further reduction in constraints by about a factor 10 as could be
obtained from the next generation instruments, would force uncomfortable modifications to current theory.

\vskip 20 pt
\centerline{  {\bf 4.  Next Generation Instruments}    }
\bigskip
\noindent
Next generation instruments in ground-based gamma-ray astronomy aim to achieve
roughly one order of magnitude improvement in sensitivity and energy threshold over present instruments.

A reduction in energy threshold has already been demonstrated by those experiments making use of existing
solar power facilities with very large reflecting surface area ($>$2000 m$^2$) to image the lateral 
distribution of EAS. Some hadronic background rejection
is available based on the expected differences in uniformity of gamma-ray and CR signatures. The 
CELESTE group have been able to achieve a $\sim$50 GeV threshold detection of the Crab 
(de Naurois 2000) and Mrk 421 (Holder 2000). The Crab has also been detected
by STACEE (Oser et al. 2001) and GRAAL (Diaz 2001) although at slightly higher energies. 
Another experiment of this type is Solar-2 (T\"umer et al. 1999). Improvements in sensitivity
will come from the addition of more mirrors, and possibly improvements in techniques to reject
background events.  

Direct sampling of EAS particles at ground-level is also possible at TeV energies. The
water \v{C}erenkov detector MILAGRO (McCullough 1999) and the air shower array in 
Tibet operating at very
high altitude have produced detections of Mrk 501 (Atkins et al. 1999, Amenomori et al. 2000) 
and Crab (Amenomori et al. 1999). 
A nice feature of these detectors is their 24 hr duty cycle and nearly 2$\pi$ sr field of view, albeit
with angular resolution and collecting area inferior to \v{C}erenkov imaging systems. Future 
developments in these areas could however provide rather sensitive all-sky monitors at sub-TeV energies.

It appears likely that the most significant improvements will come from systems using the imaging 
technique described earlier. 
H.E.S.S. (Hofmann 1999), VERITAS (Lessard 1999) and CANGAROO III (Mori 2000) will employ the 
stereoscopic imaging technique which permits an array of 
moderate-size telescopes to achieve arc-minute angular resolution. These systems consist of
arrays of $\ge 4$ telescopes each with $\sim$100 m$^2$ segmented mirrors and imaging cameras of 
$\ge$500 pixels sub-tending $\ge 3^\circ$ fields of view. MAGIC (Lorentz 1999) and MACE (Bhat 2000) 
will explore the use of very large segmented mirrors ($>$200 m$^2$) on a single telescope in 
order to push the energy 
threshold below 50 GeV, although additional telescopes may be added after experience with images
properties at energies $\le$50 GeV is gained.
It is expected that H.E.S.S., VERITAS and CANGAROO III will achieve peak gamma-ray detection rates 
(for a range of spectral indices) at energies $\sim$100 GeV or slightly lower. Spectral studies 
should be possible from this threshold energy up to $\sim$10 TeV, an unprecedented dynamic range
for this field. Table 2. summarises the salient 
features of these next generation imaging instruments, which will commence operation in the years
2002 to 2005.
\bigskip
\vbox{\tabskip=8pt 
\hrule \smallskip \hrule \smallskip
\halign to \hsize{#\hfil&#\hfil&#\hfil&#\hfil&\hfil#\hfil&#\hfil  \cr
Instrument &  Site  &  Mirror(s) & Camera       & Energy        & Epoch \cr
           &        &            & pixels       & Thres. (GeV)  & begin \cr 
           &        &            & (resol.)     &               &       \cr 
\noalign{\smallskip \hrule \smallskip} 
CANGAROO III  &  Woomera    & 4x10m   & 4x512                & 100 &  2003 \cr
              &             &         & (0.12$^\circ$)       &     &      \cr
MACE          &  India      & 1x17m   & 1x$>$800             & 10  &  2005 \cr
              &             &         & (0.1, 0.2$^\circ$)   &     &       \cr
MAGIC Phase I &  La Palma   & 1x17m   & 1x577                & 30  &  2002 \cr
              &             &         & (0.1$^\circ$)        &     &       \cr
H.E.S.S. Phase I  &  Namibia    & 4x12m   & 4x960                & 100 &  2002 \cr
              &             &         & (0.16$^\circ$)       &     &       \cr
VERITAS       &  Arizona    & 7x11m   & 7x499                & 75  &  2005  \cr
              &             &         & (0.15$^\circ$)       &     &       \cr
\noalign{\smallskip \hrule \smallskip \hrule \smallskip}}
\bigskip
{\eightpoint 
\noindent {\bf Table 2.} Summary of next generation atmospheric \v{C}erenkov imaging systems 
           approved and/or under construction. Details of each project can be found from
           references cited in the text.}
\bigskip}

Along with ground-based techniques, the development of the next space-based instruments
at gamma-ray energies GLAST (Gehrels 2000) and X-ray energies (for example XMM-Newton and 
{\it Chandra}) will advance significantly our understanding of SNRs.
Arc-second resolution now available from these X-ray satellites is allowing
detailed spatial and spectral studies of SNRs and comparisons with radio data, thereby 
improving greatly the likelihood of disentangling thermal and non-thermal X-ray components.

GLAST will sample the high energy gamma-ray sky at sensitivities and angular resolution 
an order of magnitude better than that of EGRET. A major goal of GLAST will be to 
shed light on the $>$150 unidentified EGRET sources (Hartman et al. 1999), a number 
of which are considered
SNRs (Romero et al. 1999), and also for related sources such as giant molecular clouds.
Certain questions however will not be easily accessible to GLAST. 
The origin of the Galactic CRs up to the knee will be a question best addressed by 
ground-based methods operating at higher energies since the electronic and hadronic channels described above
require multi-TeV energy particles to produce TeV gamma-rays. The huge collecting area afforded by 
ground-based techniques gives them a sensitivity advantage over satellite-based gamma-ray
systems, although
the latter of course operate with much larger fields of view ($\ge 0.5$ sr) and duty cycles.
In the context of probing spectral and morphological properties in SNRs requiring high statistics over
a broad energy range, the ground-based systems are ideal instruments.

%For extended sources such as SNRs,
%confusion with the diffuse gamma-ray background along and near the Galactic plane is also reduced at 
%ground-based energies 
%owing to its rather steeper spectral index ($E^{-2.7}$, $E\ge$10 GeV), in 
%relation to the flatter indices expected of, particularly, the hadronic channel.

\vskip 20 pt
\centerline{  {\bf 5.  Issues for Future SNR Studies at GeV/TeV Energies}    }
\bigskip
\noindent 
Summarised here are a number of issues considered important for studies
of SNRs by the next generation ground-based imaging instruments. Many of the groups building such instruments
have, or are now devoting significant effort to these with the use of Monte Carlo simulations in order 
to optimise astrophysics potential.
\bigskip

\item{1.} {\bf SNR Detectability:} The problem is that SNR will invariably be extended sources.
Included in figure 1. is an example sensitivity curve for H.E.S.S. phase I observing {\it point} sources, 
which will be more or less similar to that for VERITAS and CANGAROO-III. 
%A SNR will generally extend to radii $\sim$10 pc after having reached the 
%Sedov phase, ie. sub-tending up to $\sim 1^\circ$. 
The minimum detectable flux from an extended source $F_{\rm ext}$ of size $\sigma_{\rm src}$ may
be expressed, assuming Gaussian source morphology:\bigskip

\centerline{$ F_{\rm ext} \sim F_{\rm pt}\, \sqrt{(2\pi k \sigma^2_{\rm src} + s)/
                     (2\pi k \sigma^2_o + s)}$\hskip 80pt (1)} 
\bigskip

\item{} for $F_{\rm pt}$ the point source sensitivity, $\sigma_o$ the point spread function of the 
instrument, $s$ the number of signal counts, and $k$ the background count density (total background counts
$b=k\pi \sigma^2$), after gamma-ray selection. This equation accounts for the increase in statistical 
uncertainty of the background with 
size, contribution from the signal, and results from rearrangement of Eq. 5 in Li \& Ma 
(1983),  setting ON=$b+s$, OFF=$b$ and $\alpha=1.0$. For example a SNR of radius 
$\sigma_{\rm src} \sim 0.5^\circ$ viewed by an instrument with $\sigma_o = 0.1^\circ$, an increase in
minimum detectable flux by a factor $\sim$5 is expected for signals with $s$=200 and $b$=800 
counts\footnote{$^{\dag}$}{\eightpoint Here I have assumed $b$=800 background counts fall within the region 
$\sigma_o=0.1^\circ$ over 10 hours observation time, derived from a post-gamma-ray shape-selected trigger rate 
of 5 Hz (assuming a shape selection efficiency for background is 1\%) over a region of radius 1.5$^\circ$, 
as might be expected for H.E.S.S. and the like. Then, $s$=200 is chosen to give a $\sim 5\sigma$ 
point source significance.}.
Equation 1 neglects contributions from systematic uncertainties in estimating $b$
arising from, for example instrument performance and sky condition changes during data taking. Such errors
increase with $\sigma^2_{\rm src}$ making this potentially a dominant term if they
exceed $\sim$ few percent and the source is large. 
Instrument performance nevertheless is often able to meet this criterion via diligent screening of 
data quality.

%A second term accounting for systematic uncertainties in the signal and background,
%arising from differences in observation and instrument conditions between data taken on source
%and off source, could also enter if systematics are sufficiently large. 
%Since the systematic uncertainties will increase with $\sigma^2_{\rm src}$, this term could 
%dominate for $f$ sufficiently large.  
%Minimising $f$ involves keeping good control on instrument conditions
%and performance during the observation.

\medskip             
\item{2.} {\bf Energy Resolution and Sensitivity:} With point source sensitivities approaching 
10$^{-13}$ erg cm$^{-2}$ s$^{-1}$, and energy resolution $\le 15$\%, the spectra of sources 
with $\sim$10\% Crab flux will be determined by
H.E.S.S., VERITAS, CANGAROO-III etc. after reasonable observation time. It should therefore be possible to
discriminate the hadronic and electronic components of TeV emission, given sufficient statistics
at certain energy domains where spectral differences at maximised. 
IC and also Bremsstrahlung spectra are quite sensitive to SNR age and magnetic field values due to 
electron cooling, leading to characteristic IC 'turnovers' in spectral index which appear at 
multi-GeV to multi-TeV energies. The hadronic component will reflect more
just the upper energy limit to particle energies. Spectral differences may therefore be most
apparent at the low and high energy end of instrument sensitivities, highlighting the need to achieve a
wide dynamic range. The effect of electron cooling for example due to the high $B$ field is evident 
in figure 1.

\medskip
\item{3.} {\bf Angular Resolution:} The hadronic channel for TeV emission in SNRs will generally 
trace regions of high ambient density and regions where the SNR shock has interacted with a 
molecular cloud, leading to distinct morphological TeV features. The electronic 
channel morphology on the other hand will depend strongly on the magnetic field. For example Aharonian \& 
Atoyan (1999) argue that for SN1006, the electronic IC TeV emission would essentially fill the SNR 
under current assumptions about the magnetic field $B\le10$ $\mu$G. Instrument angular resolution
will play an important role in discriminating between TeV components. Since the angular
resolution is mildly energy-dependent (a factor $\sim$2 improvement is generally achieved at the
high energy end), it may be possible to perform such studies better at energies
well away from threshold, aligning with the arguments concerning spectral differences in point 2 above. 

\medskip
\item{4.} {\bf Field of View and Acceptance} SNR sizes generally exceed comfortably the instrument 
point spread function. A wide radius ($\ge 1.5^\circ$) achieving a roughly flat gamma-ray acceptance 
is therefore desirable. This aspect is also very important in searches for the EGRET unidentified sources,
where positional uncertainties of $\le 1^\circ$ are noted.
 
\vskip 20 pt
\centerline{  {\bf 6.  Conclusions}    }
\bigskip
\noindent

Over the past decade, ground-based observations of SNRs at TeV gamma-ray energies have been carried 
out in an attempt to establish them as the primary sites of Galactic CR's.  
With the exceptions of three cases, studies of SNRs at TeV energies have revealed 
non-detections with upper limits at levels of the order 10\% Crab. Such limits do constrain models
on CR production in SNRs, but generally lie close to the  conservative edge of the 
rather large parameter space available. The next generation ground-based instruments employing the
imaging atmospheric \v{C}erenkov technique from $\sim$50 GeV to 10 TeV are expected however in
conjunction with new X-ray satellites and the forthcoming space-based gamma-ray instrument GLAST, to sample 
the sky at sufficient sensitivity and resolution to provide serious constraints on the theory that 
SNRs are responsible for Galactic CRs. 
Negative results from SNR observations with these ground-based instruments, would confront us
with a number of not necessarily exclusive consequences, assuming that we retain the 
diffusive shock acceleration framework: 
(1) Limiting the energy content in accelerated particles, particularly hadrons, in
SNRs to $E_{\rm cr}\le 10^{49}$ erg; (2) CR source spectra generally steeper than that predicted 
from current theory, (3) our understanding of SNR dynamics is clearly lacking, (4) and/or that alternative 
sources contribute significantly to CR acceleration. For the last point Galactic sources such as 
pulsar-driven nebulae or plerions and 
microquasars/X-ray binaries, all of which appear capable of maintaining particle acceleration at 
sufficient luminosity for very long times, may be alternatives. Definite answers concerning  
these type of questions should be available not long after 2002, when the first of the next-generation
instruments is fully online.
  
\bigskip
{\eightpoint 
\noindent {\bf Acknowledgements} I would like to thank the organising committee of the GAME 2001 
           workshop for the invitation, and gratefully acknowledge receipt of a von Humboldt Fellowship.
           F.A. Aharonian, D. Horns and H.J. V\"olk are thanked for helpful comments.}
\bigskip

% Sample references
\bigskip
\centerline{\bf References}
\bigskip
{\eightpoint\parindent=0pt\everypar={\hangindent=0.5 cm}
% References in the format of the Bulletin of the Astronomical Society of India
% using 8 pt fonts
% leave one line blank between two references to force a paragraph break
%scussions.

 Aharonian F.A., et al. 2001a, A\&A 370, 112  % Cas-A paper

 Aharonian F.A., et al. 2001b, A\&A in press  % Tycho paper

 Aharonian F.A., Atoyan A.M. 1999, A\&A 351, 330

 Akhperjanian A., Sahakian V. 1999 Astropart. Phys. 12, 157

 Atkins R., Benbow W., Berley D. et al., 1999 ApJ 525, L25

 Atoyan A.M., Aharonian F.A., Tuffs, R.J., V\"olk H.J. 2000, A\&A 355, 211

 Amenomori M., Ayabe S., Cao P.Y. et al., 1999 ApJ 525, L93

 Amenomori M., Ayabe S., Cao P.Y. et al., 2000 ApJ 532, 302
   
 Barrau A., Bazer-Bachi R., Beyer E. et al., 1998 Nucl. Inst. Meth. A 416, 278

 Baring M.G., Ellison D.C., Reynolds S.P. et al., 1999 ApJ, 523, 311

 Berezhko E.C., V\"olk H.J. 1997 Astropart. Phys. 7, 183

 Berezhko E.C., V\"olk H.J. 2000 A\&A 357, 283

 Bhat C.L. 2000 ``High Energy Gamma-Ray Astronomy'', AIP Conf. Proc. 558, 582

 Blandford R., Eichler D. 1987 Phys. Rep. 154, 1

 Buckley J.H., Akerlof C.W., Carter-Lewis D.A. et al., 1998 A\&A 329, 639

 Daum A., Hermann G., Hess M., et al. 1997 Astropart. Phys. 8, 1

 Diaz M. 2001 Proc. XXXVIth Renc. de Moriond, Les Arcs, France, in press

 de Naurois M. 2000 ``High Energy Gamma-Ray Astronomy'', AIP Conf. Proc. 558, 540

 Drury L.O'C., Aharonian F.A., V\"olk H.J. 1994, A\&A 287, 959

 Drury L.O'C., Kirk J., Duffy P. 2000 ``GeV/TeV Gamma-Ray Astrophysics Workshop'', AIP Conf. Proc. 
     515, 183

 Drury L.O'C., Ellison D.C., Aharonian F.A., Berezhko A., et al. 2001 ``ISSI Workshop on Astrophysics
            of Cosmic Rays'', astro-ph/0106046

 Esposito J., Hunter S.D., Kanbach G., et al., 1996 ApJ 461, 820

 Fegan S. 2001 In proc. of ``The Nature of Unidentified Galactic High Energy Gamma Ray Sources'', Puebla, Mexico, astro-ph/0102324

 Finley J.P., Bond I.H., Bradbury S.M., et al. 1999 ``GeV/TeV Gamma-Ray Astrophysics Workshop'', AIP 
      Conf. Proc. 515, 301 

 Gaisser T., Protheroe R.J., Stanev T. 1998 ApJ 492, 219

 Gehrels N. ``High Energy Gamma-Ray Astronomy'', AIP Conf. Proc. 558, 3

 Goret P., Gouiffes C., Nuss E., et al. 1999 Proc. 26th ICRC (Salt Lake City) 3, 496

 Green D.A. 2000 ``A Catalogue of Galactic Supernova Remnants (2000 August version)'', Mullard Radio 
     Astronomy Observatory, Cavendish Laboratory, Cambridge, United Kingdom  
        (http://www.mrao.cam.ac.uk/surveys/snrs/). 

 Hartman R.C. et al. 1999 ApJS 123, 79

 Hillas A.M. 1985, Proc. 19th Int. Cos. Ray Conf. (La Jolla) 3, 445
 
 Hofmann W., Jung I., Konopelko A., et al. 1999 Astropart. Phys. 12, 135

 Hofmann W. 1999 ``GeV/TeV Gamma-Ray Astrophysics Workshop'', AIP Conf. Proc. 515, 500

 Holder J. 2000 ``High Energy Gamma-Ray Astronomy'', AIP Conf. Proc. 558, 635

 Kirk J.G., Dendy R.O. 2001, J.Phys. G: Nucl.Part.Phys.  in press 

 Koyama K., Petre R.,  Gotthelf E.V., et al. 1995 Nature 378, 255

 Koyama K., Kinugasa K., Matzusaki K., et al. 1997 PASJ 49, L7 

 Laming J.M. 1998 ApJ 499, 309L

 Laming J.M. 2001 ApJ 546, 1149L

 LeBohec S., Degrange B., Punch M. et al. 2000 Nucl. Inst. Meth. A 416, 425

 Lessard R.W., Bond I.H., Boyle P.J., et al., 1999 Proc. 26th ICRC (Salt Lake City) 3, 488 {\it OG.2.2.16}

 Lessard R.W., 1999 Astropart. Phys. 11, 243

 Li T., Ma Y. 1983 ApJ 272, 317

 Lorentz E. 1999 ``GeV/TeV Gamma-Ray Astrophysics Workshop'', AIP Conf. Proc. 515, 510

 Lucareli F., Konopelko A., Rowell G., et al. 2000 ``High Energy Gamma-Ray Astronomy'', AIP Conf. Proc. 558, 779

 McCullough J.F. 1999, Proc. 26th Int. Cos. Ray Conf. (Salt Lake City) 2, 369 

 Mori M., 2000 ``High Energy Gamma-Ray Astronomy'', AIP Conf. Proc. 558, 578

 Muraishi H., Tanimori T., Yanagita S., et al. 2000 A\&A 354, L57

 Naito T., Takahara F. 1994 J.Phys.G 20, 477

 Oser S., Bhattacharya D., Boone L.M. et al., 2001 ApJ 547, 949 

 Romero G.E., Benaglia P., Torres D.F. 1999 A\&A 348, 868

 Rowell G.P., Naito T., Dazeley S.A., et al. 2000 A\&A 359, 337

 Slane P., 2001 astro-ph/0104348    %X-ray SNR review

 Slane P., Hughes J.P., Edgar R.J., et al. 2001 ApJ accepted (astro-ph/0010510)

 Sturner S.J., et al. 1997 ApJ 490, 619

 Tanimori T., Hayami Y., Kamei S., et al. 1998 ApJ 497, L25

 T\"umer T., Bhattacharya D., Mohideen U., et al. 1999 Astropart. Phys. 11, 271

 V\"olk H.J. 2001 Proc. XXXVIth Renc. de Moriond, Les Arcs, France (astro-ph/0105356)

 Yoshikoshi T., Dazeley S.A., Gunji  S., et al., 1999 Astropart. Phys. 11, 267

 Weekes T.C., Cawley M.F., Fegan D.J., et al. 1989 ApJ 342, 379

 Weekes T.C. 2000 ``High Energy Gamma-Ray Astronomy'', AIP Conf. Proc. 558, 15

% End of section heading
%\endref
}                                         % End of references
% leave one line blank before the closing braces.
\end